# Biopolymers: life's mechanical scaffolds

Federica Burla[†], Yuval Mulla[†], Bart E. Vos, Anders Aufderhorst-Roberts, Gijsje H. Koenderink[*]

AMOLF, Department of Living Matter, Biological Soft Matter group, Science Park 104, 1098 XG Amsterdam, the Netherlands
[*]Corresponding author: g.koenderink@amolf.nl
[†] These authors contributed equally to this work.

**Abstract | The cells and tissues that make up our body juggle contradictory mechanical demands. It is crucial for their survival to be able to withstand large mechanical loads, but it is equally crucial for them to produce forces and actively change shape during biological processes such as tissue growth and repair. The mechanics of cell and tissues is determined by scaffolds of protein polymers known as the cytoskeleton and the extracellular matrix, respectively. Experiments on model systems reconstituted from purified components combined with polymer physics concepts have already successfully uncovered some of the mechanisms that underlie the paradoxical mechanics of living matter. Initial work focussed on explaining universal features such as the nonlinear elasticity of cells and tissues in terms of polymer network models. However, living matter exhibits many advanced mechanical functionalities that are not captured by these coarse-grained theories. In this Review, we focus on recent experimental and theoretical insights revealing how their porous structure, structural hierarchy, transient crosslinking, and mechanochemical activity confer resilience combined with the ability to adapt and self-heal. These physical insights improve our understanding of cell and tissue biology and also provide a source of inspiration for synthetic life-like materials.**

From a physicist's perspective, cells and tissues are fascinating materials because they combine an extraordinary mechanical strength with the ability to grow, reshape and adapt to environmental conditions. This paradoxical combination of strength and dynamics is essential for supporting life. Mechanical strength is crucial because cells and tissues constantly experience large mechanical loads[1]. With every breath we take, endothelial cells lining blood vessels and epithelial cells in the lung for instance experience large tensile stresses. With every step we take, muscles and tendons stretch while cartilage compresses. Cells and tissues are able to cope with these mechanical challenges because they are supported by filamentous protein networks that provide an efficient means of mechanical scaffolding. Unlike man-made polymers, however, biopolymer networks not only provide mechanical support, but they also actively reconfigure themselves. Cells are able to actively adjust their stiffness in response to environmental conditions and produce forces that drive cell division and motility. At the tissue level, cellular force generation drives the formation of tissues and organs in developing embryos and the regeneration of tissues in adult organisms.

      Cells are mechanically supported by the *cytoskeleton*, a composite network of three types of protein filaments: actin filaments, microtubules, and intermediate filaments (Fig. 1)[2]. It is generally believed that intermediate filaments are particularly important for the protection of cells against large deformations, since they form resilient and long-lived elastic networks. By contrast, actin filaments and microtubules form dynamic networks that actively generate forces with the aid of motor proteins and proteins that regulate filament (de)polymerization. Connective tissues such as skin and arteries are instead supported by the *extracellular matrix*, which is likewise a composite network made up of polymers with complementary physical properties[3]. Collagen forms a rigid fibrillar network that endows tissues with a high tensile strength, whereas proteoglycans and glycosaminoglycans form a soft hydrogel that holds water and confers resistance against

compressive loads. Connective tissues furthermore contain varying amounts of the elastomeric protein elastin and other fibrous proteins such as fibronectin and laminin, which regulate cellular functions.

Cells adhere to the extracellular matrix through transmembrane proteins known as integrins, which directly bind components of the extracellular matrix such as collagen and fibronectin and indirectly couple to the actin and intermediate cytoskeleton through accessory proteins[4]. They thus transfer contractile forces generated by the actin cytoskeleton to the extracellular matrix. Cells therefore actively remodel and tense the extracellular matrix, in a process contributing to tissue formation and wound healing. Conversely, the architecture and mechanical properties of the matrix strongly influence cell behavior. Cells probe the physical properties of the matrix through the contractile forces they apply at integrin adhesions (*mechanosensing*) and they convert this mechanical information into biochemical signals that elicit a cellular decision such as cell growth and differentiation (*mechanotransduction*). In the past decade it has become well-established that mechanical forces steer many biological processes that are physiologically important such as wound healing, but also pathological processes such as cancer metastasis[5]. This realization has driven the emergence of mechanobiology as a new research field and has given a strong boost to the field of cell and tissue biophysics.

There are two fundamentally different approaches one can take to investigate the physical basis of cell and tissue mechanics. The first approach is top-down and involves mechanical measurements and phenomenological modelling of whole cells or tissues. Such measurements have revealed that living matter exhibits surprisingly universal mechanics. First, cells behave as viscoelastic materials with a power-law dependence of the storage (elastic) and loss (viscous) shear moduli on the deformation frequency, suggesting that they dissipate elastic stresses with a broad spectrum of relaxation times[6]. Second, cells and tissues exhibit a nonlinear elastic response to mechanical loading. They often strain-stiffen, but depending on the rate, amplitude and type of loading (i.e. compression, shear, or tension) they can also soften[7-9]. Third, cells and tissues are usually under substantial internal stress. The contractile activity of cells generates stress in the cytoskeleton, which is transferred to the extracellular matrix through integrin adhesions[10, 11]. Due to their charged nature, proteoglycans in the extracellular matrix can generate additional mechanical stress[12]. Unfortunately, understanding the physical mechanisms that underlie these intriguing collective mechanical properties is extremely challenging due to the molecular and structural complexity of living systems and the presence of mechanochemical feedback. This complexity has motivated a second, bottom-up approach to cell and tissue physics. In this approach, components of the cytoskeleton and/or the extracellular matrix are purified and studied in isolation or together with a limited set of regulatory proteins. This reductionist approach has succesfully driven the development of quantitative theoretical frameworks to describe cell and tissue mechanics and biological processes such as cell migration [13, 14]. Current models usually coarse-grain biopolymers as elastic beams or semiflexible polymers, motivated by their large size (10-100 nm diameter) and high bending rigidity compared to standard synthetic polymers. However, recently there has been a growing realization that biopolymers exhibit many material properties that are not captured by these simple models.

Here we review recent advances in the field of cell and tissue biophysics, focusing on reductionist studies of the mechanics of their biopolymeric scaffolds. After a brief summary of the elastic properties of biopolymer networks, we highlight recently discovered mechanical functionalities that arise from the unique biomolecular make-up of living matter. In particular we discuss the interplay of the polymer network and the background solvent, the mechanical synergy that arises from combining multiple components with distinct properties, the resilience provided by the structural hierarchy of biopolymer, the role of transient crosslinking, the physical mechanisms and biological relevance of plasticity, and finally the role of active driving.

**Elastic properties of biopolymer networks**
Cytoskeletal and extracellular polymers are supramolecular filaments with a complex and highly organized molecular structure dictated by specific interactions between the constituent proteins. Examples are the double-helical architecture of actin filaments and the quarter-staggered packing structure of collagen (Fig. 1). Cytoskeletal filament assembly is driven by reversible noncovalent interactions. This dynamic assembly is integral to the biological functions of the cytoskeleton, where actin filaments and microtubules often need to (dis)assemble rapidly in response to biochemical or mechanical signals. In contrast, extracellular matrix polymers such as collagen are more stable due to covalent crosslinks created by enzymes[15].

Mechanical models of cytoskeletal and extracellular matrix polymers usually coarse-grain the filaments by a smooth linear rod that resists bending with a modulus $\kappa$ and stretching with a modulus $\mu$. At finite temperatures, thermal fluctuations cause the filaments to bend as a function of their persistence length $l_p$, defined as the decay length of angular correlations along the polymer contour. The persistence length is related to the bending modulus as $\kappa = k_B T l_p$, where $k_B$ is Boltzmann's constant and T is temperature. Biopolymers are categorized on the basis of the ratio between $l_p$ and the contour length L as flexible ($l_p \ll L$), semiflexible ($l_p \sim L$), or stiff ($l_p \gg L$). Collagen and microtubules have persistence lengths in the mm-range and are therefore examples of stiff filaments, whereas actin filaments and intermediate filaments have persistence lengths in the µm-range and are therefore semiflexible[16-18]. An example of a flexible biopolymer is hyaluronan, a polysaccharide in the extracellular matrix with a ~4-8 nm persistence length[19].

Biopolymers are assembled in load-bearing networks by a variety of mechanisms. The simplest mechanism is by entanglements that naturally arise from steric interactions (Fig. 2a). At high enough densities, polymers constrain each other's motions to snake-like paths along their contour, as conceptualized by the reptation model[20, 21]. The micrometer-sized length of cytoskeletal filaments has made it possible to directly observe filament reptation by fluorescence microscopy[22]. Entangled biopolymer solutions can only store elastic energy on short time scales, because at longer time scales the filaments escape the constraints posed by entanglements[23]. Long-term mechanical stability is therefore only possible in the presence of long-lived filament interactions, which can occur by branching or crosslinking (Fig. 2b). In the cytoskeleton, actin filaments and microtubules are branched and crosslinked by a large set of specialized proteins[24, 25], while intermediate filaments are crosslinked through a combination of accessory proteins and cation-mediated interactions[26]. The transient nature of these filament connections turns cytoskeletal networks into viscoelastic materials. By contrast, the extracellular matrix has a more elastic character due to covalent crosslinking. For example, the collagen framework is covalently crosslinked by lysyl oxidase[15]. When polymerized on its own, purified collagen tends to form networks through a combination of branching and crosslinking[27, 28], while in the body, collagen assembly and mechanics is tightly regulated in a tissue-specific manner by cells and accessory matrix molecules[29].

Measurements on reconstituted biopolymer networks have revealed a general tendency to stress-stiffen in response to shear or uniaxial tensile loads and to stress-soften under compressive loads[30-33] (Fig. 3a). Theoretical modeling has shown that these nonlinear elastic properties are an intrinsic feature of filamentous networks. Compression-induced network softening involves a competition between softening due to polymer buckling and stiffening due to polymer densification upon solvent efflux[31-34]. Much more is known about the stiffening response upon tensile or shear loading. Interestingly, the mechanisms that govern stiffening are fundamentally different for semiflexible and rigid polymer networks. Semiflexible polymer networks stiffen because the conformational entropy of the polymers is reduced as they are pulled taut along the direction of principal strain[35] (Fig. 3b). The elastic modulus can be calculated by averaging over the entropic force-extension response of the constituent filaments[30], provided that the network is densely crosslinked so that it deforms uniformly (affine) down to length scales on the order of the crosslink distance[36]. The elastic modulus is expected to increase with applied (shear) stress according to a power law with an exponent of 3/2, which is indeed observed in case of actin and intermediate filaments[37, 38]. The onset strain where stiffening sets in is governed by the amount of excess length stored in thermal fluctuations of polymer segments between adjacent crosslinks, and is therefore a function of the persistence length and crosslink density. Networks of actin and intermediate filaments are highly strain-sensitive because stiffening already sets in at strains of just a few percent and the stiffness can easily increase by a factor 10-100 before rupture. This strain-sensitivity is believed to mechanically protect cells by preventing large deformations. Moreover, it allows cells to tune their stiffness by molecular motor activity, as explained later on. Given these advantages, there is a growing interest in mimicking strain-sensitivity in synthetic polymer gels. Although synthetic polymers are typically flexible[30], recently several groups have for the first time successfully created synthetic polymers that are sufficiently stiff to exhibit strain sensitivity[39-41].

Networks of stiff (athermal) filaments such as collagen also strain-stiffen, but in this case the nonlinearity is an emergent phenomenon that arises at the network level (Fig. 3c). This form of non-linearity is related to the network connectivity. Since biopolymers form networks through a combination of branching and crosslinking, the average coordination number ranges between 3 and 4.[27, 28] We refer to these networks as subisostatic because the coordination number is below the Maxwell criterion of 6 required for mechanical stability of networks of springs[42]. Unlike springs, however, fibres can form stable subisostatic networks because of their large bending rigidity[43]. Filamentous networks are soft at small strains because they deform in a nonaffine manner dominated by fibre bending[44, 45]. However, shear or tensile strains drive a transition to a rigid state dominated by fibre stretching because the fibres align along the principal direction of strain. This transition occurs at a critical strain set by the network connectivity[28, 45, 46]. Collagen networks are highly strain-sensitive given that nonlinearity usually sets in already at strains of ~10% and the stiffness can increase by 100-fold before network rupture. Strain-stiffening is thought to help prevent tissue rupture by preventing high strains and to promote long-range mechanical communication between cells[47].

**Poroelastic effects in biopolymer networks**

Biopolymer networks are biphasic systems since they combine a solid porous phase comprised of protein fibers with a fluid phase that typically takes up more than 95% of the total volume. Compressive or tensile deformations that change the volume of the system will necessarily induce fluid flow through the network due to the incompressibility of water (Fig. 4a). This causes a time-dependent mechanical response that is referred to as *poroelasticity*[48]. When the deformation is fast, the system will respond like an incompressible material because the load is supported primarily by the incompressibility of the interstitial fluid[49]. By contrast, the system responds like a compressible material when the deformation is slow enough to allow for fluid outflow (in case of compression) or inflow (in case of extension). The typical time scale $\tau$ for a fluid of viscosity $\eta$ to flow across a distance $d$ through a polymer network with pore size $\xi$ and shear modulus $G$ can be estimated using a two-fluid model for a linearly elastic polymer network in a viscous background fluid[50, 51], according to $\tau \sim \eta\, d^2 / G\, k$. Here $k \sim \xi^2$ is the network's hydraulic permeability.

Poroelastic effects are well-known in the context of tissue biomechanics but were long thought to be unimportant inside the cells because of their small size (5-20 µm). However, a seminal study on blebbing cells showed that poroelastic effects actually do impact cell mechanics on time scales relevant to cell motility[52]. When the cell membrane locally detaches from the cytoskeleton, spherical membrane protrusions called blebs are formed. Active contraction of the actin-myosin cortex creates a compressive stress that initially only locally increases the hydrostatic pressure, whereupon fluid flow inflates the detached membrane. Pressure equilibration across the cell takes on the order of ~10 s because of the small mesh size of the cytoskeleton (~10 nm) and the high viscosity of the cytoplasm[53]. Later, AFM nanoindentation and microrheology measurements confirmed these findings[53, 54]. Cells may exploit the slow equilibration of hydrostatic pressure to generate blebs or lamellipodial protrusions for locomotion[55].

It was recently discovered that poroelasticity also significantly impacts the shear rheology of biopolymer networks, even though shear deformations are volume-conserving as opposed to compressive and tensile deformations (Fig. 4b). Sheared polymer networks develop a normal force perpendicular to the direction of shear, which tends to be negative (contractile) for semiflexible and rigid biopolymers and positive (extensile) for flexible polymers[56], an effect known as the Poynting effect[57]. If one neglects the influence of the interstitial fluid, the normal force from the Poynting effect is always calculated to be negative because network segments that develop tension outnumber nodes under compression for networks of springs[58], semiflexible polymers[56] and subisostatic networks of rigid fibres[45, 46]. However, in the presence of a fluid phase, the normal stress is always positive at short time scales because of the strong viscous coupling between the polymer network and the interstitial fluid. The normal stress switches in sign from positive to negative at time scales corresponding to the characteristic time for fluid flow introduced above[51]. Therefore this time scale is highly sensitive to biopolymer rigidity such that rigid polymers such as collagen switch sign in normal force almost instantaneously because the micrometer-sized pores enable rapid equilibration of hydrostatic pressure. Conversely, flexible polymer hydrogels exhibit negative normal stress only after many hours because their pore size is in the nanometer range and the gel is effectively an incompressible material on experimentally accessible time scales.

**Mechanical synergy in multicomponent biopolymer networks**
A striking feature of the cytoskeleton and the extracellular matrix is that both are composite mixtures of biopolymers with different mechanical and dynamic properties. The synergy between polymers that individually already have rich properties allows Nature to access a wide range of mechanical properties to meet the requirements of different cell and tissue types despite using the same building blocks. Cartilage, for instance, needs to simultaneously resist tensile and compressive loads, and achieves this through the interplay between a fibrous collagen network and a proteoglycan meshwork[59]. Migrating cells need to combine resilience with directional motion through fluidisation, and rely therefore on coupling between actin, intermediate filaments and microtubules[60]. Composite biopolymer networks have only recently begun to be investigated by quantitative rheological measurements and theoretical modelling. The focus thus far has been on two-component systems, but even this simplified context already creates an enormously rich parameter space where the network mechanics can be tuned by variations in the persistence lengths of the two polymers, their relative and absolute densities, and the interconnectivity among the two components (Fig. 5).

In theoretical studies, this complex phase space has been mainly explored in the limit of permanently crosslinked networks that juxtapose rigid and (semi)flexible polymers. When both polymers form percolating networks, the linear elastic modulus of the composite can become substantially larger than the sum of the moduli of the separate networks[61]. In such systems, the biopolymer with a lower rigidity forms a denser elastic background due to its smaller mesh size, which in turn increases the effective bending rigidity of the more rigid biopolymer[61, 62]. Such a synergistic increase in the linear modulus has been experimentally observed in composites of actin and the intermediate filament protein vimentin, which differ in persistence length by a factor 10 (with $l_p$ = 10 µm and 1 µm, respectively)[63], although this was not confirmed in a more recent study, perhaps due to subtle differences in the filament interactions[64]. Networks of (semi)flexible polymers have also been predicted to reinforce rigid polymers against compressive loads[62], an effect that has indeed been observed in actin-microtubule composites[65] and is thought to be important for cells crawling through soft matrices[66]. In the context of the extracellular matrix, collagen-hyaluronan composites were also reported to exhibit an enhanced resistance to compressive loading compared to collagen alone[67]. However, in this case the mechanism was not elastic, but viscous in origin: hyaluronan enhances the viscosity of the fluid in the interstices of the collagen matrix and thus increases the hydraulic resistance to fluid outflow. Since glycosaminoglycans tend to swell in hypotonic solutions, they can also induce prestress when interpenetrated with a collagen network[68], which can change the nonlinear elastic response of collagen due to its stress-sensitivity[69].

Surprisingly, mechanical enhancement can also be achieved for composites in which only one of the two polymers forms a percolating network. In this case the dominant component determines the linear elastic modulus, while the inclusions influence the nonlinear elastic response[70]. Rigid polymer inclusions are expected to lower the threshold shear strain required to induce strain-stiffening of semiflexible polymers by making the strain field more affine[70-73]. This effect has indeed been confirmed experimentally for composite networks of actin and microtubules[74, 75]. Furthermore, rigid polymer inclusions are predicted to induce compressibility in an otherwise almost incompressible matrix, because they constrain the displacement field[76], a phenomenon observed in co-entangled actin and microtubule composites[77].

An important challenge in experimental studies of composite networks is that the constituent polymers can influence each other's organization through steric constraints, direct interactions, or depletion effects. Structural changes caused by such mutual interactions have for instance been reported for composites of actin and intermediate filaments[78, 79] and collagen and glycosaminoglycan composites[80]. It will be important in future studies to gain better control over the network structure of composites through the assembly kinetics and the use of bifunctional crosslinking agents such as plectins and spectraplakins[60]. An alternative approach is to create hybrids of biopolymers and synthetic polymers or fully synthetic hybrid networks, which provide better control over the interaction partners and assembly conditions of the polymers[81, 82]. Furthermore, the theoretical predictions of the relation between the stress and strain field in composite networks have yet to be examined experimentally, for example by confocal rheometry[36].

In the cytoskeleton, the crosslinks that connect the filaments are proteins that in some cases directly influence the network mechanics by contributing their own compliance. An extreme example is filamin, a V-shaped protein whose two-actin binding domains are connected by long and flexible linker domains. Filamin drastically changes the nonlinear elastic response of actin networks, from the 3/2 power-law stiffening observed with rigid crosslinks such as α-actinin to an approximately linear stiffening response[83]. This effect has been explained by modelling actin-filamin networks as composites of rigid filaments and wormlike chain crosslinkers[84]. Compliant crosslinks or combinations of crosslinkers with different rigidities thus provide additional control knobs to tune the nonlinear mechanics of cytoskeletal networks[85-87].

**Mechanical strength enhancement by structural hierarchy**

When protein biopolymer networks are subjected to large (>50%) strains, they will break unless the constituent polymers are able to elongate. Recent studies have shown that several cytoskeletal and extracellular protein biopolymers are extremely extensible because their molecular packing structure can change under strain (Fig. 6). One mechanism for filament elongation is by sliding of protein subunits relative to one another. Subunit sliding has been observed for microtubules and for collagen fibers, which are both bundles of thin protofilaments associated by lateral interactions that are weaker than the longitudinal interactions[88, 89]. Although the bending stiffness of both filament types is length-dependent due to protofilament sliding[88, 89], the filaments are rather inextensible and already break at strains of 50-80%[16, 90, 91]. Bundling of actin filaments with crowding agents or crosslink protein generates filamentous structures that can lengthen, giving rise to rate-dependent force-extension behavior[92, 93].

An alternative mechanism for filament elongation is by molecular unfolding of the protein subunits. This phenomenon is well-documented for intermediate filaments, which can be stretched to more than 3 times their rest length using the tip of an atomic force microscope[18, 94]. Spectroscopic measurements of the secondary structure content and X-ray scattering measurements of the molecular packing structure showed that stretching is mediated by a conformational transition of the protein subunits from alpha-helical to beta-sheet[95, 96], which sets in at tensile strains of about 10%. As a result, the mechanical response of the filaments is strongly dependent on loading rate[18]. A similar α-helix to β-sheet transition has been proposed as an explanation for the remarkable extensibility of the fibres formed by the blood clotting protein fibrin based on X-ray scattering and spectroscopy measurements on fibrin networks[97-99] and single molecule evidence for unfolding[100, 101]. However, this mechanism has not yet been definitively proven because the complex architecture of fibrin fibres also provides alternative mechanisms for elongation. The fibres are thick bundles of ~100 protofibrils that are interconnected by long linker domains that are highly flexible because they are largely unstructured[102, 103]. Several studies suggested that linker stretching can account for the extreme extensibility of single fibrin fibres without the need to invoke unfolding of the structured domains[104, 105]. It could well be that both mechanisms act in unison[106]. In a conceptually similar manner, the elastin filaments that confer resilience to skin, lung and vascular tissues combine long disordered protein domains that are flexible and extensible with ordered domains that confer rigidity and tensile strength [107-109]. Due to the large number of organizational levels of biopolymers, it is still a large challenge to dissect the precise molecular mechanisms that orchestrate their elastomeric properties. Multi-technique approaches that correlate the mechanical response measured at the fibre or network level with molecular changes as measured through small-angle X-ray scattering[97, 99] or vibrational spectroscopy[98] are needed, coupled to multiscale modeling that connects molecular models to fiber and network models through systematic coarse-graining[110].

The extensibility of intermediate filaments, fibrin, and elastin enable cells and tissues to cope with large mechanical strains. Moreover, these filaments nonlinearly stiffen as they are stretched, which has been predicted to enhance their flaw tolerance[111]. Both of these features would be highly desirable in the design of synthetic tissues. Unfortunately, it is still difficult to realize the hierarchical structure that is characteristic of protein biopolymers in fully synthetic materials. Current efforts to make bioinspired resilient materials therefore mainly use either natural or designed recombinant proteins as building blocks[81, 112-114]. DNA nanotechnology offers another promising route towards hierarchical materials[115].

**Time-dependent rheology due to transient crosslinking**
Until now we have only considered the elastic properties of biopolymer networks. However, cells are actually viscoelastic materials with time-dependent mechanical properties, since the linker proteins that mediate cytoskeletal crosslinking only bind transiently[116]. Crosslinker dynamics are crucial for cell functions such as migration, division and morphogenesis, because they allows cells to dynamically remodel their interior and change shape[117, 118].

The mechanical consequences of transient crosslinking have so far mainly been studied in the context of actin networks. At the single-molecule level, actin crosslinkers have typical bond lifetimes of several seconds[119, 120]. At the network level, this translates in elastic behavior at time scales shorter than the bond lifetime and viscoelastic flow on longer timescales[121]. Interestingly, this viscoelastic flow does not follow a simple Maxwell model with a single relaxation time, but instead follows power law behavior characteristic of multiple relaxation times[122] (Fig. 7). In the linear elastic regime, both the storage and loss modulus show an $\omega^{1/2}$ dependence. Even though there is only a single microscopic time scale for cross-linker unbinding, there is a broad spectrum of macroscopic relaxation times since each filament is crosslinked to the surrounding networks by many crosslink proteins. Stress relaxation therefore requires many independent binding and unbinding events[122, 123]. In the nonlinear regime, the network response becomes dependent on time as well as stress because some crosslinker proteins exhibit *slip bond* behavior, meaning that they dissociate faster in the presence of an applied force[119]. As a consequence, actin networks soften at small loading rates due to forced crosslink unbinding, whereas they stiffen due to nonlinear elasticity when the loading rate exceeds the crosslinker unbinding rate[124-126]. Intriguingly, several linkers, including α-actinin, filamin and vinculin, exhibit an opposite response to loading known as *catch bond* behavior, whereby the bond lifetime initially increases with force because loading exposes a hidden binding site[127-129]. Catch bonds have indeed been shown to delay the onset of relaxation and flow in actin networks[130].

A complication in studying reconstituted actin networks is that the structure is often kinetically controlled due to dynamic arrest during the polymerization process as the growing filaments get entangled and crosslinked[131, 132]. Kinetic trapping can cause the presence of long-lived internal stresses that take many hours to relax because of the slow dynamics of crosslinker-governed network relaxation[133, 134]. It is unclear whether dynamic arrest is relevant in the context of cells, where actin filaments are constantly disassembled and nucleated anew.

The extracellular matrix has a more elastic character than cells because the collagen framework is covalently crosslinked[15]. However, studies on reconstituted collagen networks showed that stress relaxation is significantly enhanced under strain, due to force-dependent unbinding of the bonds holding together the fibers[135]. Furthermore, the interstitial space of collagen networks in tissues is filled with a soft hydrogel background comprised of hyaluronic acid and other transiently crosslinked components, introducing additional mechanisms for stress relaxation[136]. It will be interesting to investigate the collective dynamics that result from the composite architecture of the matrix, especially since recent work has revealed that the viscous response of the matrix, in addition to rigidity, has a significant impact on the behavior and function of cells[137, 138].

**Plasticity, fracturing and self-healing**
Upon cyclic loading, cytoskeletal networks exhibit plasticity, or mechano-memory. Plasticity arises because mechanical loading causes dissociation of the crosslinkers and the dissociated crosslinkers can diffuse and rebind elsewhere[116]. Crosslink redistribution can freeze in shear-induced fiber alignment, causing network hardening[139, 140]. When the shear stress is too high, actin networks completely lose mechanical percolation. Experimentally, the rupture strength is known to depend on the actin filament length and crosslink density[141] and on the microscopic properties of the crosslinkers, including their compliance[83, 85]. The microscopic mechanism of rupture is still poorly understood. We recently showed by one-dimensional modelling of bond arrays that dynamic crosslink unbinding should make transient networks inherently prone to fracturing, as local fluctuations in crosslinker density propagate into large-scale cracks[142, 143]. Although cytoskeletal networks are prone to fracture, they are also inherently self-healing. Broken crosslinks are capable of re-forming[144] and the filaments themselves can even self-repair by the addition of new monomers[145, 146]. In cells, the nucleation and growth of new filaments can further promote self-healing[147]. The self-healing potential of transiently connected networks has already been picked up in materials science as highlighted by several exciting recent examples of self-healing synthetic polymers[148, 149].

Extracellular matrix networks including collagen and fibrin also exhibit plasticity upon cyclic loading, but in this case the fibers themselves form new bonds in the deformed state[32, 150]. Once the external stress is released, these new bonds are stretched, causing the build-up of internal contractile stress that nonlinearly stiffens the network. Because of the complex molecular packing structure of the fibers, additional plasticity can arise at the level of the fibers themselves[151]. In the case of non-crosslinked networks of collagen or fibrin, cyclic shearing has been observed to cause fiber lengthening, presumably through subunit sliding, causing a delayed onset of strain-stiffening[152]. There is a growing recognition that these mechano-memory effects are relevant for normal tissue development but also for pathological processes such as fibrosis and cancer progression. By exerting contractile forces, cells irreversibly remodel the extracellular matrix and generate rigid, aligned fiber tracts[151, 153, 154]. These rigid tracts in turn promote cellular force generation through positive mechano-chemical feedback.

**Active material properties**
A unique feature of the cytoskeleton is that it is turned into an active material by molecular motors, which convert chemical energy provided by ATP hydrolysis into mechanical work[155]. Motors take advantage of the structural polarity of actin filaments and microtubules that results from the head-to-tail assembly of the protein subunits to step unidirectionally along these filaments. The material properties of the actin cytoskeleton are mostly governed by non-muscle myosin-II motors[7, 156]. Individually, myosin-II motors cannot generate contractile stress because they are non-processive, meaning that they are only bound to actin for a small fraction of the ATP hydrolysis cycle. Stress generation requires myosin-II assembly into bipolar filaments of ~10-30 motors. Since the motor domains are exposed on the two ends, bipolar myosin filaments can slide anti-parallel actin filaments along each other (Fig. 8a). In the absence of crosslinks, this sliding activity can fluidize actin networks by relieving entanglement constraints[157], which perhaps explains myosin-driven softening observed for suspended cells[158]. In the presence of crosslinks, myosin-driven sliding instead causes contractile stress build-up[159]. In principle, extension should be equally likely as contraction. However, several mechanisms bias actin-myosin networks towards contraction[160]. An important contribution seems to come from the asymmetric response of crosslinked fibrous networks to compressive versus tensile strain[161, 162]. Simulations showed that collective fiber buckling in the vicinity of a local contractile force center will always rectify the stress towards strongly amplified isotropic contraction in disordered networks[162]. This principle applies equally well to extracellular matrix networks containing embedded contractile cells (Fig. 8b). Active gel models furthermore predict that contractile stress will stiffen filamentous networks because of their nonlinear elastic response to stress[163-165] (Fig. 8c). Indeed, motor-driven stiffening has been experimentally confirmed for actin/myosin-II networks[166, 167] as well as for fibrin and collagen networks with cells[47, 168].

Another important source of activity in the cytoskeleton is the constant turnover of actin filaments and microtubules that is driven by nucleotide hydrolysis by the filaments themselves[169]. Hydrolysis of ATP (in case of actin) and GTP (in case of microtubules) allows the filaments to polymerize on one end while depolymerizing on the other end[169, 170]. Filament turnover is expected to dissipate motor-driven stress because tensed filaments are removed by depolymerization while new filaments are produced in a stress-free state[171-173]. A recent experimental study indeed confirmed that filament treadmilling speeds up stress relaxation in actin networks[174]. Experiments on cell extracts showed that the combination of motor activity and actin turnover leads to multiple dynamic steady states including long-range flow patterns[175]. Given the complexity of extracts, which contain many thousands of distinct proteins[176], it will be interesting to test these findings also in reconstituted networks.

The active material properties of cytoskeletal networks have already inspired several exciting synthetic realizations, such as synthetic polymer networks driven by fuel-dependent polymer tread-milling[177] or light-driven molecular rotors[178] and DNA-based networks driven by processive enzymes[179].

**Conclusions**

A defining feature of living matter is the combination of two contradictory mechanical functionalities: the capacity to resist substantial loads and the ability to actively change its shape, architecture and mechanics. The understanding of the design principles underlying these functionalities has been made possible by quantitative experiments on reconstituted biopolymers coupled with theoretical modelling. This reductionist approach has revealed that living matter owes its mechanical strength to the fibrous architecture of the cytoskeleton and extracellular matrix, the hierarchical structure of the fibers, the presence of active internal forces, and the synergistic combination of different biopolymers in composite networks.

Despite the numerous advances in our understanding of biopolymer mechanics, many open questions remain. Arguably the most challenging of these is how living systems maintain mechanical strength while actively deforming. This is especially difficult to understand in the context of cells, because cell deformability requires transient crosslinking, but transient bond dynamics makes materials vulnerable to rupture. We speculate that catch bond crosslinkers may help cells to circumvent this problem, since they tend to accumulate in stressed regions[180]. A further factor is the complementarity of the three cytoskeletal systems, which have traditionally been regarded as independent with separate cellular tasks. However, there is mounting evidence that they function in a coupled manner through interactions mediated by crosslink and motor proteins and shared signalling pathways[60]. Microtubules and actin stress fibers for instance align and polarize intermediate filaments, while aligned intermediate filament structures in turn serve as a long-lived template that guides microtubule growth[181]. Intermediate filaments also integrate the contractile forces generated by actin across the cell[182]. We anticipate that studies of reconstituted composite cytoskeletal networks will provide a powerful strategy to elucidate the collective active and passive material properties that emerge from cytoskeletal teamwork. Future experimental progress will be aided by advanced techniques developed for measuring mechanics *in situ*, such as optical microrheology and molecular force sensors[183], while progress in modelling will benefit from advances in coarse-grained approaches and statistical frameworks to describe active matter[110, 184, 155].

Even though connective tissues are often regarded as much more static structures than cells, everyone who has recovered from a broken bone or has performed body building knows that bones and muscles adapt to mechanical loading. In fact the architecture of our bones is precisely optimized for the local loading conditions in the body. The dynamics that mediate this adaptivity are driven by cells, which constantly synthesize collagen and other extracellular matrix constituents and degrade the matrix by secreting proteolytic enzymes[185]. There is intriguing evidence that collagen displays a use-it-or-lose-it functionality: collagen fibrils under high strain are protected from enzymatic degradation, whereas fibrils under small strain are enzymatically destroyed [186]. As a result, collagenous materials dynamically adapt to physiological loads, selectively strengthening and pruning themselves to retain a structure in the principal loading direction. Finding the mechanisms that lead to this counterintuitive behaviour would be helpful in understanding pathologies such as fibrosis and would guide the design of materials for tissue regeneration.

Understanding the mechanical design principles of living matter is important to understand the mechanistic basis of diseases associated with genetic defects in cytoskeletal and matrix proteins such as skin fragility and heart muscle failure[187, 188]. Furthermore, living matter has come to be regarded as a paradigmatic example of a growing class of soft condensed matter known as *active matter*[189]. Studies of reconstituted systems are providing an instructive road map for the creation of biomimetic materials with life-like features. It remains a challenge to realize the active driving and hierarchical structuring that is unique to living matter, and therefore hybrid materials combining synthetic and biological building blocks (proteins or even cells) provide a promising avenue.

**Acknowledgements**
We thank Kristina Ganzinger for critically reading the manuscript and Cristina Martinez Torres for help in acquiring the confocal image shown in figure 1. We thank Fred MacKintosh (Rice University & VU Amsterdam) and Chase Broedersz (TUM) for many stimulating discussions about many of the topics covered in this review. We furthermore gratefully acknowledge financial support by the European Research Council (Starting Grant 335672-MINICELL) and by the Industrial Partnership Programme Hybrid Soft Materials, which is carried out under an agreement between Unilever Research and Development B.V. and the Netherlands Organisation for Scientific Research (NWO).

**Contributions**
All authors contributed to writing this paper.

**Highlighted references**
1) Block, J. et al. *Nonlinear Loading-Rate-Dependent Force Response of Individual Vimentin Intermediate Filaments to Applied Strain*. Physical Review Letters 118, 048101 (2017)
   **Pioneering experimental study of the stress-strain response of single intermediate filaments, demonstrating that monomer unfolding explains their remarkable**

     **extensibility.**
2) Sharma, A. et al. *Strain-controlled criticality governs the nonlinear mechanics of fibre networks.* Nature Physics 12, 584-587 (2016).
**Combined experimental and theoretical study showing that the nonlinear elasticity of fibrous networks such as collagen reflects a strain-controlled continuous phase transition.**
3) de Cagny, H.C. et al. *Porosity Governs Normal Stresses in Polymer Gels.* Physical Review Letters 117, 217802 (2016).
**Rheology experiments combined with theory reveal that sheared polymer networks universally switch from positive to negative normal stress on a time scale governed by the network porosity, explaining the opposite behaviors encountered in synthetic and biopolymer gels.**
4) Moeendarbary, E. et al. *The cytoplasm of living cells behaves as a poroelastic material.* Nature materials 12, 253-261 (2013).
**Direct evidence of the importance of poroelasticity in cell rheology based on microindentation and microrheology experiments.**
5) Brangwynne, C.P. et al. *Microtubules can bear enhanced compressive loads in living cells because of lateral reinforcement.* The Journal of cell biology 173, 733-741 (2006).
**Combined in vivo and in vitro study showing the importance of the composite architecture of the cytoskeleton for cell mechanics.**
6) Brown, A.E., Litvinov, R.I., Discher, D.E., Purohit, P.K. & Weisel, J.W. *Multiscale mechanics of fibrin polymer: gel stretching with protein unfolding and loss of water.* Science (New York, N.Y.) 325, 741-744 (2009).
**Impressive multi-scale experimental study of the role of structural hierarchy in enhancing the mechanical strength of fibrin.**
7) Bonakdar, N. et al. *Mechanical plasticity of cells.* Nature materials 15, 1090-1094 (2016).
8) Mulla, Y., Oliveri, G., Overvelde, J.T.B. & Koenderink, G.H. *Crack Initiation in Viscoelastic Materials.* Physical Review Letters 120, 268002 (2018).
**Theoretical modelling reveals that transient networks are susceptible to spontaneous crack initiation with a critical crack length that can be predicted from the nonlinear local bond dynamics.**
9) Munster, S. et al. *Strain history dependence of the nonlinear stress response of fibrin and collagen networks.* Proceedings of the National Academy of Sciences of the United States of America 110, 12197-12202 (2013)
**Confocal imaging directly reveals that fibrin and collagen networks exhibit plasticity due to a combination of network- and fibre-level remodeling in response to cyclic shear.**
10) Koenderink, G.H. et al. *An active biopolymer network controlled by molecular motors.* Proceedings of the National Academy of Sciences of the United States of America 106, 15192-15197 (2009).
**Experimental demonstration that myosin motor filaments actively stiffen actin networks by generating internal stress, revealing the mechanism by which myosin contractility controls cell stiffness.**

136. Lou, J., Stowers, R., Nam, S., Xia, Y. & Chaudhuri, O. Stress relaxing hyaluronic acid-collagen hydrogels promote cell spreading, fiber remodeling, and focal adhesion formation in 3D cell culture. *Biomaterials* **154**, 213-222 (2018).
137. Gong, Z. *et al.* Matching material and cellular timescales maximizes cell spreading on viscoelastic substrates. *Proceedings of the National Academy of Sciences of the United States of America* **115**, E2686-e2695 (2018).
138. Bennett, M. *et al.* Molecular clutch drives cell response to surface viscosity. *Proceedings of the National Academy of Sciences of the United States of America* **115**, 1192-1197 (2018).
139. Schmoller, K.M., Fernandez, P., Arevalo, R.C., Blair, D.L. & Bausch, A.R. Cyclic hardening in bundled actin networks. *Nature communications* **1**, 134 (2010).
140. Majumdar, S., Foucard, L.C., Levine, A.J. & Gardel, M.L. Mechanical hysteresis in actin networks. *Soft Matter* **14**, 2052-2058 (2018).
141. Kasza, K.E. *et al.* Actin filament length tunes elasticity of flexibly cross-linked actin networks. *Biophysical journal* **99**, 1091-1100 (2010).
142. Mulla, Y., Oliveri, G., Overvelde, J.T.B. & Koenderink, G.H. Crack Initiation in Viscoelastic Materials. *Physical review letters* **120**, 268002 (2018).
143. Mulla, Y. & Koenderink, G.H. Crosslinker mobility weakens transient polymer networks. *arXiv:1805.12431* (2018).
144. Sanchez, T., Chen, D.T., DeCamp, S.J., Heymann, M. & Dogic, Z. Spontaneous motion in hierarchically assembled active matter. *Nature* **491**, 431-434 (2012).
145. Schaedel, L. *et al.* Microtubules self-repair in response to mechanical stress. *Nature materials* **14**, 1156-1163 (2015).
146. Noding, B., Herrmann, H. & Koster, S. Direct observation of subunit exchange along mature vimentin intermediate filaments. *Biophysical journal* **107**, 2923-2931 (2014).
147. Sano, K. *et al.* Self-repairing filamentous actin hydrogel with hierarchical structure. *Biomacromolecules* **12**, 4173-4177 (2011).
148. Yang, Y. & Urban, M.W. Self-healing polymeric materials. *Chemical Society reviews* **42**, 7446-7467 (2013).
149. Yanagisawa, Y., Nan, Y., Okuro, K. & Aida, T. Mechanically robust, readily repairable polymers via tailored noncovalent cross-linking. *Science (New York, N.Y.)* **359**, 72-76 (2018).
150. Kurniawan, N.A. *et al.* Fibrin Networks Support Recurring Mechanical Loads by Adapting their Structure across Multiple Scales. *Biophysical journal* **111**, 1026-1034 (2016).
151. Ban, E. *et al.* Mechanisms of Plastic Deformation in Collagen Networks Induced by Cellular Forces. *Biophysical journal* **114**, 450-461 (2018).
152. Munster, S. *et al.* Strain history dependence of the nonlinear stress response of fibrin and collagen networks. *Proceedings of the National Academy of Sciences of the United States of America* **110**, 12197-12202 (2013).
153. Kim, J. *et al.* Stress-induced plasticity of dynamic collagen networks. *Nature communications* **8**, 842 (2017).
154. Hall, M.S. *et al.* Fibrous nonlinear elasticity enables positive mechanical feedback between cells and ECMs. *Proceedings of the National Academy of Sciences of the United States of America* **113**, 14043-14048 (2016).
155. Gnesotto, F.S., Mura, F., Gladrow, J. & Broedersz, C.P. Broken detailed balance and non-equilibrium dynamics in living systems: a review. *Reports on progress in physics. Physical Society (Great Britain)* **81**, 066601 (2018).
156. Kollmannsberger, P., Mierke, C.T. & Fabry, B. Nonlinear viscoelasticity of adherent cells is controlled by cytoskeletal tension. *Soft Matter* **7**, 3127-3132 (2011).
157. Humphrey, D., Duggan, C., Saha, D., Smith, D. & Kas, J. Active fluidization of polymer networks through molecular motors. *Nature* **416**, 413-416 (2002).

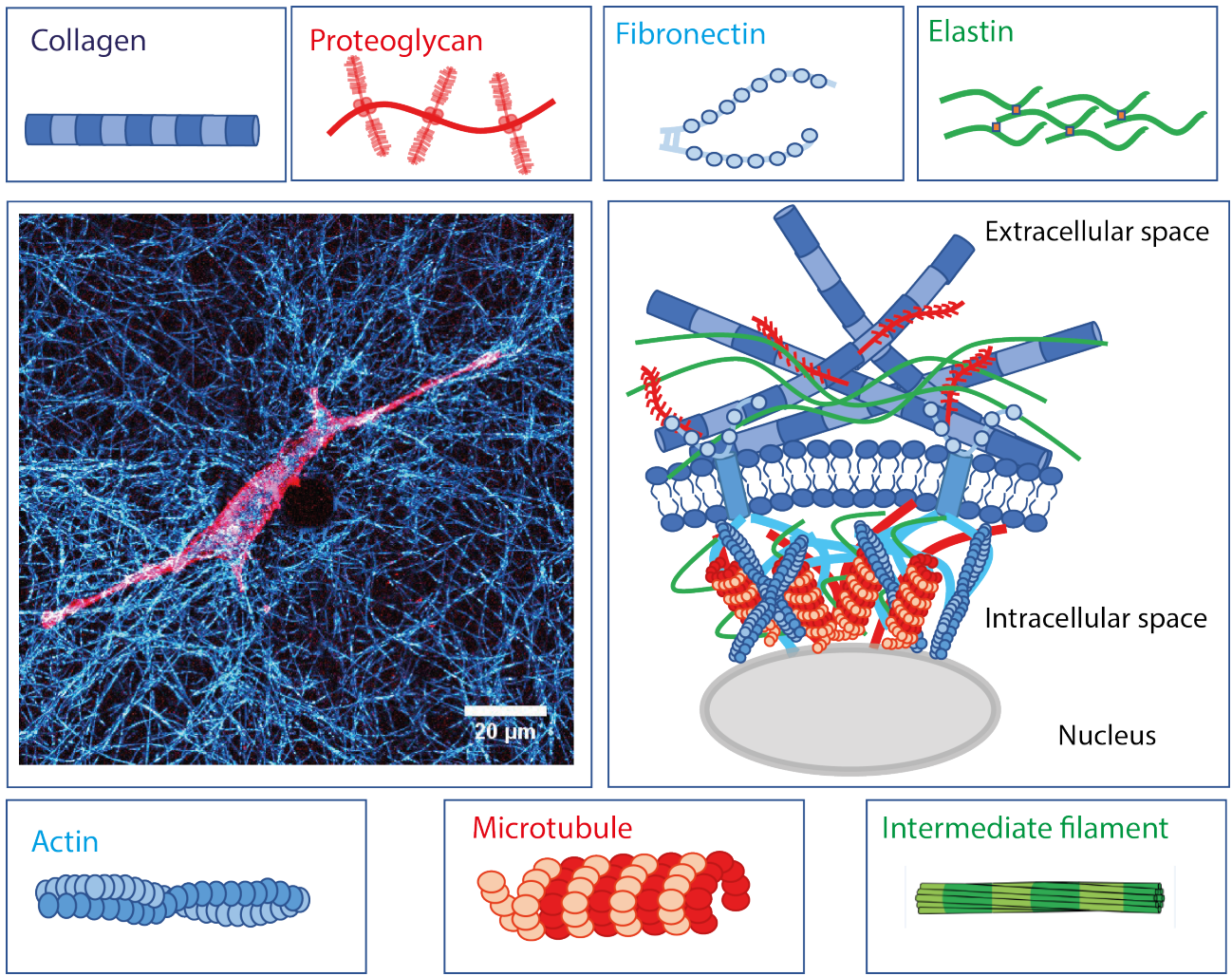

**Fig. 1: Cells and tissues are mechanically supported by biopolymer networks.** The central panel shows a confocal microscopy image of a cell (actin cytoskeleton labeled in red) adhered to a collagen matrix (blue fibers) together with a schematic view of the cytoskeleton and extracellular matrix connected across the cell membrane via integrin adhesion proteins. Note that the extracellular matrix *in vivo* is three-dimensional in some tissues such as skin, while it forms a two-dimensional sheet in other tissues such as epithelia. The upper panel shows the most prevalent biopolymers present in the extracellular matrix, while the lower panel shows the three filaments that make up the cytoskeleton of the cell.

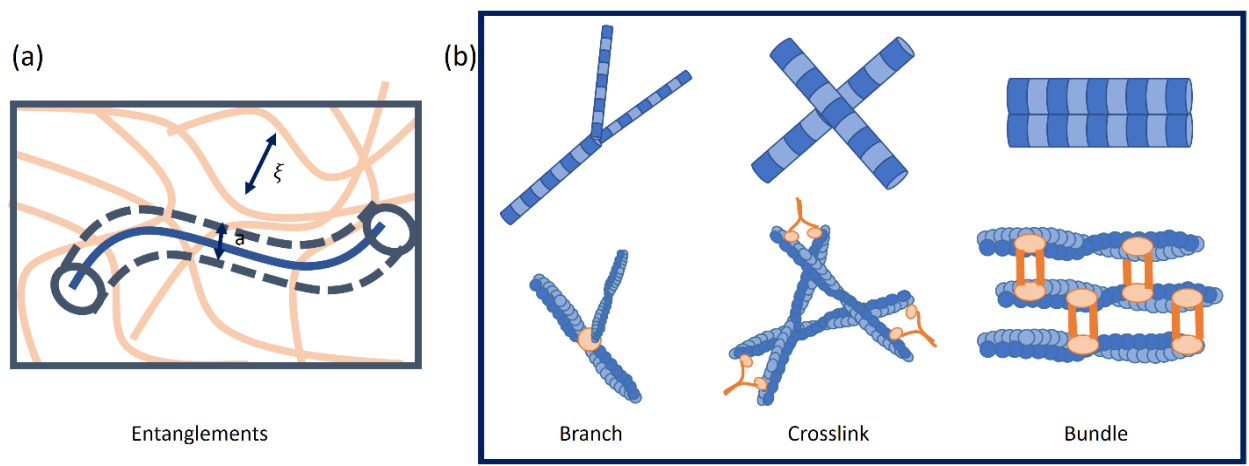

**Fig. 2: Biopolymers form networks via multiple mechanisms**. (a) Biopolymers entangle when their density is high enough such that they sterically hinder each other's transverse motion. The dashed cylinder indicates the snake-like path along which each polymer is forced to reptate. The arrows indicate the tube width *a* and network mesh size $\xi$. (b) Branches, crosslinks and bundles can be formed either by intermolecular interactions of the filaments themselves, as in the case of collagen (top), or by accessory proteins, as in the case of actin (bottom).

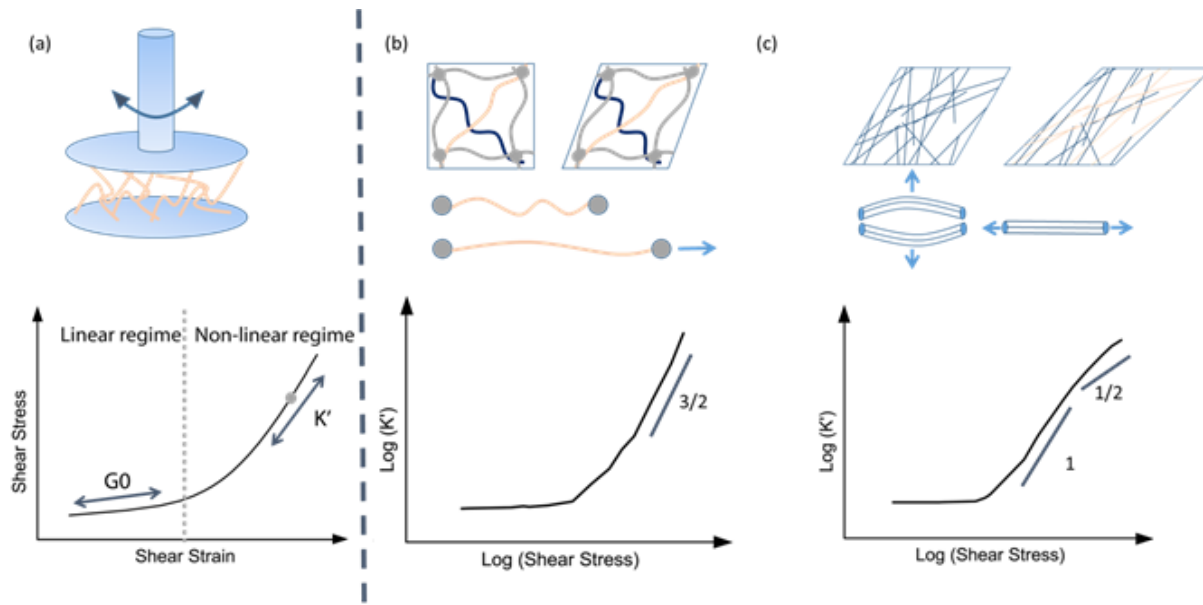

**Fig. 3: Nonlinear elasticity in biopolymer networks.** (a) The nonlinear elastic response of biopolymer networks can be probed by subjecting networks polymerized between two plates to an oscillatory or steady shear deformation. The stress/strain response is linear at small strains, where the slope gives the linear modulus $G_0$, but curves up at high strain, where the slope gives the differential modulus *K'*. (b) Semiflexible polymer networks strain-stiffen due to the entropic resistance of the thermally undulating filaments against stretching, giving rise to a characteristic 3/2 power-law stiffening. (c) Stiff polymer networks strain-stiffen by undergoing a transition from a soft, bending-dominated state to a stiff, stretching-dominated state, giving rise to a power-law stiffening regime with an exponent close to 1 at moderate stress and a ½ stress-stiffening at high stress.

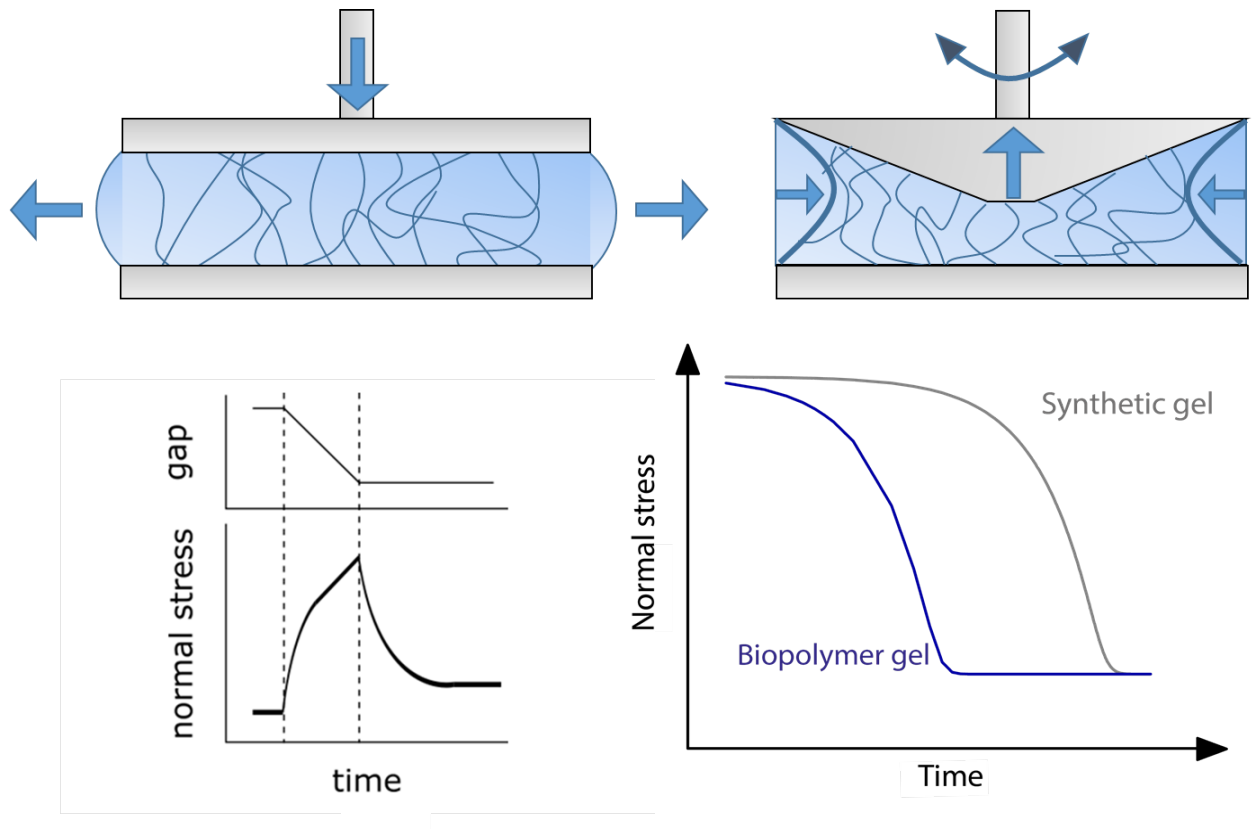

**Fig. 4: Poroelasticity of biopolymer networks.** (a) Upon compression, fluid is squeezed out of the network causing a time-dependent normal force along the axial direction. (b) Upon shearing by rotation of the upper cone, hydrostatic pressure is built up, which relaxes by an inward, radial contraction of the network relative to the solvent (blue). This results in an exponential decay of the normal stress as a function of time after the application of a constant shear stress at $t = 0$, with a time constant that is set by the pore size and therefore tends to much smaller for biopolymer gels than synthetic gels.

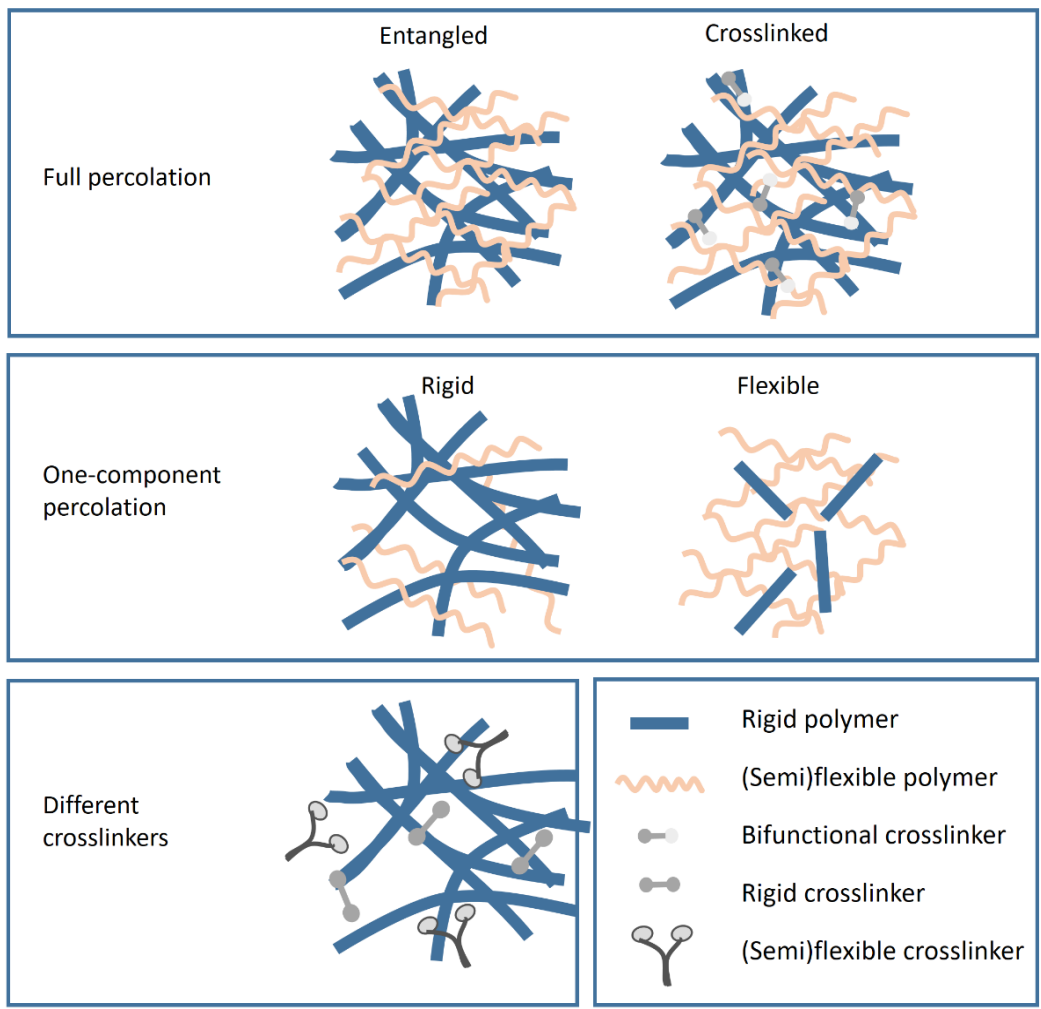

**Fig. 5: Different types of composite networks with an enhanced mechanical response.** Composites can exist of rigid and (semi)flexible polymers (top and middle), or from one polymer crosslinked with a combination of rigid and flexible linker proteins (bottom).

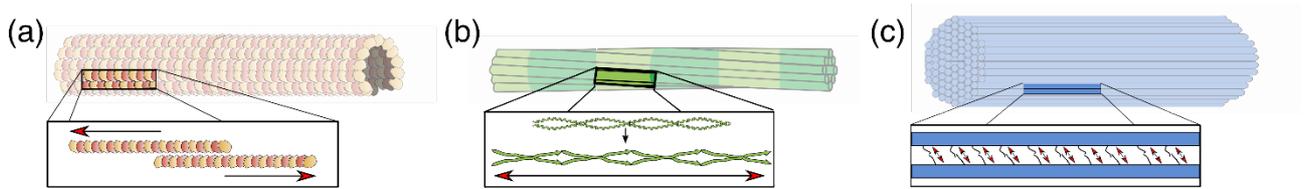

**Figure 6: The hierarchical assembly of biopolymers introduces several mechanisms for elongation.** These include sliding of subunits as observed with microtubules (a), forced unfolding of protein subunits, as observed with intermediate filaments (b), and stretching of disordered, flexible linkers that connect subunits, as observed with fibrin fibers (c).

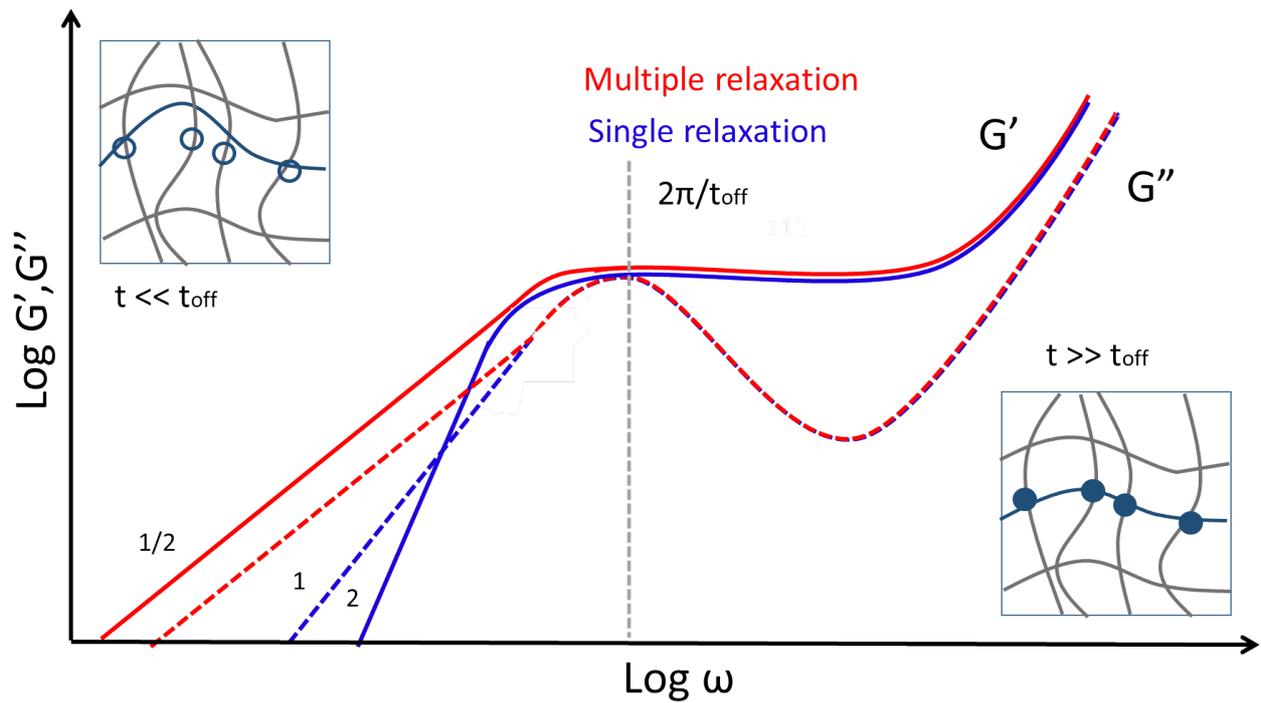

**Figure 7. Time-dependent response of polymer networks crosslinked by linkers that unbind at a rate $1/\tau_{off}$ to an oscillatory shear strain.** Flexible polymer networks behave as Maxwell fluids that undergo a transition from elastic to fluid behavior at a single characteristic frequency $\omega_{off}$. Instead, semiflexible polymer networks exhibit a broad distribution of relaxation times at frequencies below $\omega_{off}$ because stress relaxation requires many independent linker binding and unbinding events. Note that the loss and storage moduli increase at high frequencies due to viscous drag that hampers filament fluctuations.

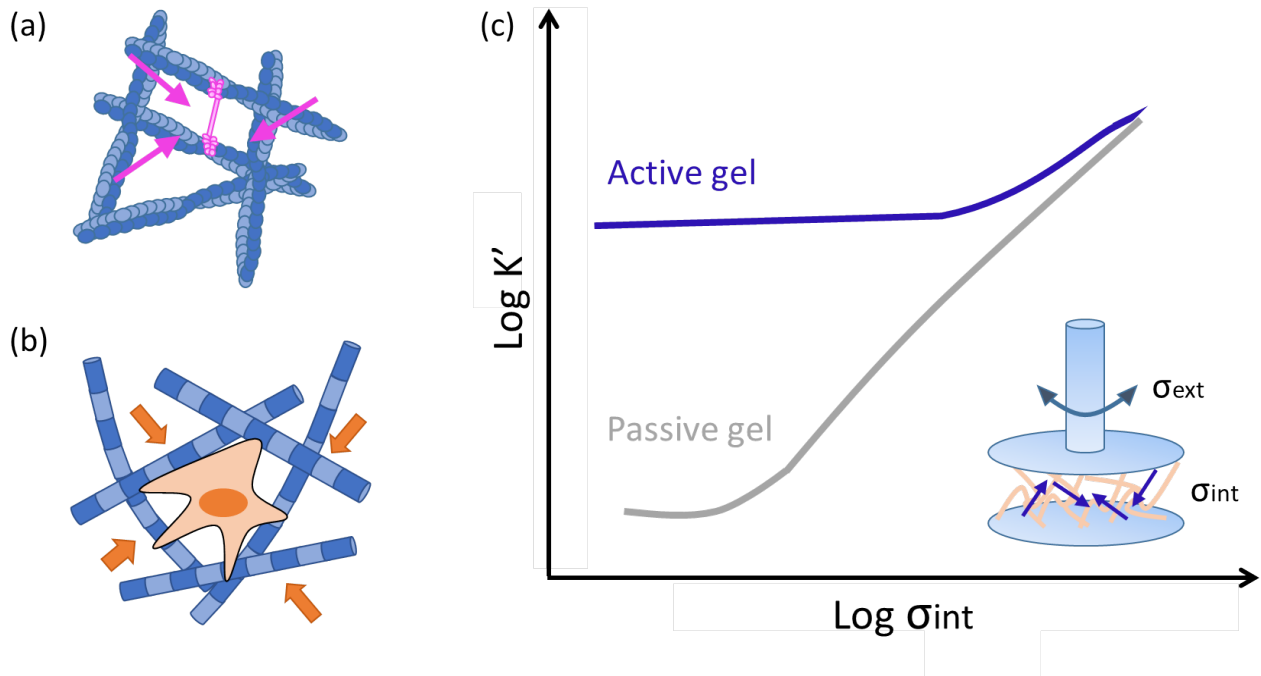

**Figure 8. Active control over biopolymer network mechanics by contractility.** (a) Myosin motors form bipolar filaments (green) that contract cytoskeletal actin networks (purple). (b) Cells contract the extracellular matrix by transferring contractile forces generated by actin and myosin through focal adhesions. (c) Active contraction makes cytoskeletal and extracellular matrix networks stiffer than their passive (equilibrium) counterparts, because the network elasticity responds nonlinearly to internal stress.